# Energy Efficient IoT Virtualization Framework with Peer to Peer Networking and Processing

Zaineb T. Al-Azez, Ahmed Q. Lawey, Taisir E.H. El-Gorashi, Jaafar M.H. Elmirghani

*Abstract*— In this paper, an energy efficient IoT virtualization framework with P2P networking and edge computing is proposed. The proposed network encompasses IoT objects and relay devices. In this network, the IoT task processing requests are served by peers. The peers in our work are represented by IoT objects and relays that host virtual machines (VMs). We have considered three scenarios to investigate the saving in power consumption and the system capabilities in terms of task processing. The first scenario is the relays only scenario, where the task requests are processed using relays only. The second scenario is the objects only scenario, where the task requests are processed using the IoT objects only. The last scenario is a hybrid scenario, where the task requests are processed using both IoT objects and VMs. We have developed a mixed integer linear programming (MILP) model to maximize the number of processing tasks served by the system and minimize the total power consumed by the IoT network. We investigated our framework under the impact of VMs placement constraints, fairness constraints between the objects, tasks number limitations, uplink and downlink limited capacities, and processing capability limitations. Based on the MILP model principles, we developed an energy efficient virtualized IoT P2P networks heuristic (EEVIPN). The heuristic results were comparable to those of the MILP in terms of energy efficiency and tasks processing. Our results show that the hybrid scenario serves up to 77% (57% on average) processing task requests, but with higher energy consumption compared to the other scenarios. The relays only scenario can serve 74% (57% on average) of the processing task requests with 8% saving in power consumption compared to the hybrid scenario. In contrast, 28% (22% on average) of task requests can be successfully handled by applying the objects only scenario with up to 62% power saving compared to the hybrid scenario. The results also revealed the low percentage of addressed task requests in the objects only scenario resulting from the capacity limits of the IoT objects' processors. In addition, the small difference between the serving percentage of hybrid scenario and relays only scenario resulted from the allowed internal processing of objects in the hybrid scenario.

## I. INTRODUCTION

THE dramatic recent developments in IoT were mainly driven by the tremendous need and benefits that can be gained from connecting our physical world to the Internet. It is expected that there will be 50 billion (and by some estimates, more) IoT interconnected devices in the coming years [1]. This growth in the number of connected devices opens the doors to new applications, for example in agriculture, transportation, manufacturing, smart homes, smart healthcare, and M2M communications [2], [3]. Many challenges such as energy efficiency, reliability, security, interoperability and scalability have to be overcome before the planned growth in the number and functionalities of IoT can be realized [4]. Given the expected number of devices, one of the most important challenges is energy efficiency and hence greening the associated networks, which grabbed attention in both the academic and industrial domains. Cloud computing is investigated as one of the solutions to the energy efficiency challenge in networks and data centers [5]-[8]. However, with the large data generated by the connected IoT objects (expected to generate 2.3 trillion gigabytes of data every day by year 2020) [2], emerging cloud computing with IoT poses new challenges which have to be addressed. Among these challenges is the hunger for more processing capabilities, high communication bandwidth, security, and latency requirements [9].

A number of solutions were suggested to address these issues. The work started with distributed content placement, thus bringing content closer to users [10], distributed data centers, thus bringing the processing capabilities closer to users and IoT devices [11] and distributed processing of big data, where processing the huge data generated by IoT devices near the source can extract knowledge from the data and hence transmit the small volume 'extracted knowledge' messages, thus saving network and processing resources and hence energy [12]. However, a different and potentially more efficient solution, advocated here, is to process the IoT data by the IoT objects themselves or by the devices in the nearest layer to these objects. According to Allied Business Intelligence (ABI), it is expected that 90% of the data created by the endpoints will be processed and stored locally rather than being handled by the conventional clouds [10]. Since some complicated data processing tasks cannot be done by most of the IoT devices and sensors because of their limited capabilities, edge computing is proposed to provide more resources to serve such tasks in efficient and fast ways. One of the suggested ways to do this is the dynamic installation of virtual machines (VMs) in the edge cloud to process the raw data generated by the tasks requested by the IoT objects. The processed results are then sent back to the objects [2].

Manuscript received xxxx; revised xxxx; accepted xxxx. Date of publication xxxx; date of current version xxxx. This work is funded from the Engineering and Physical Sciences Research Council (EPSRC) for the INTERNET (EP/H040536/1) and STAR (EP/K016873/1) projects. Zaineb T. Al-Azez,

Ahmed Q. Lawey, Taisir E.H. El-Gorashi, Jaafar M.H. Elmirghani are with the School of Electronic and Electrical Engineering, University of Leeds, Leeds, LS2 9JT, U.K. (e-mail: elztaa@leeds.ac.uk; A.Q.Lawey@leeds.ac.uk; j.m.h.elmirghani@leeds.ac.uk).

In [13], we considered a single IoT network consisting of IoT network elements (relays, coordinators and gateways). In [13], data processing and traffic aggregation were done by VMs hosted in cloudlets, where these mini clouds are distributed over the IoT network elements. The work was extended in [14] where two separated IoT networks were considered with the deployment of a Passive Optical Access Network (PON). The main goal of our previous work was to investigate the potential energy efficiency gains that can be made if use is made of distributed cloudlets at the edge of the network compared to centralized cloudlets at highest layer of the implemented model. There is a recent trend in research toward proposing IoT platforms based on local computing close to the objects such as fog and edge computing. Such platforms have many common characteristics with our proposed architecture. In [15], a combination of fog computing and microgrid is proposed in order to reduce the energy consumed by IoT applications. A set of measurements and experiments were implemented considering different processing and traffic requirements. In [15], dynamic decisions can be made by the proposed IoT gateway to minimize the consumed energy by choosing the most efficient location for processing a task in the fog or in the cloud. This decision is affected by the type of deployed IoT application, weather forecasting and the availability of renewable sources. An edge computation platform is presented in [16] where the design of an IoT gateway virtualized environment for IoT applications is proposed using lightweight virtualization technologies. In this work [16], IoT data processing can be achieved by making use of container-based virtualization technologies such as Docker containers.

IoT devices can make use of P2P communication capabilities and architectures [17]. A number of advantages could be reaized by using P2P communication systems compared to conventional communication systems such as energy efficiency, traffic reduction [17] and reliability. Based on the potential energy efficiency advantage, we introduce our energy efficient IoT network considering a combination of P2P communication between the IoT objects and edge computing while installing VMs in the relays. Computing tasks and the communication between the peers in our network is achieved through two stages, in the first stage, objects send the requests for tasks to be served by other peers (represented by IoT objects and relays hosting VMs) through the directly connected relays in the network. In the second stage the results of the processed tasks are received. We assume the traffic generated by task requests is reduced after processing by different percentages depending on the complexity of the requested tasks.

The remainder of this paper is organized as follows: In Section II, we describe the MILP model developed to optimize the network and hence construct an energy efficient P2P IoT network. Section III discusses the MILP model results. In Section IV, we introduce our network heuristic and discuss its results. Conclusions are given in Section V.

## II. ENERGY EFFICIENT MILP FOR P2P IOT NETWORKS

The MILP model developed considers the architecture shown in Fig. 1. The proposed architecture is constructed of two layers. The first layer represents the IoT objects. The upper layer consists of the relay devices that realize traffic transportation between peers. In our framework, each object is capable of processing three types of tasks that are required by other objects. The task processing capabilities and task requirements for the IoT objects are specified by the MILP model parameters. Each relay node has the ability to host VMs in order to process the tasks requested by IoT objects. The number of relays that can handle all task types is limited to a subset of total number of relays. For example, in the results section we consider a scenario in which 10 out of 25 relays host VMs that can handle all tasks types.

Fig. 1 illustrates all the processing cases we have considered in our P2P platform. Internal processing is shown in case (a), where the object has the ability to process its own request. Consequently, the network power consumption associated with sending the task request to another object or relay or receiving a task result from them will be eliminated. One application of this case might be in smart lights. In case (b), the object sends its task request to the object's neighbor (the directly connected relay device) to be processed by the hosted VM, for example a healthcare device. Some of the objects in our model have the ability to process task requests generated by other objects but considering fairness constraint limitations. The fairness constraint states that each object should reciprocate equally to other objects choosing it to process its requested task. Object to object communication such as two Arduino devices with different capabilities is illustrated in case (c). The last task processing case is case (d). In this case, none of the objects themselves or the other objects or even the VM hosted by its directly connected relay have the ability to process the requested tasks. In spite of that, relays can process all types of tasks, but the capacity of each relay-processor is limited to a specific maximum workload. So, in order to process this task, the relay sends the task request to other relays to be processed by the nearest possible relay hosting VM (keep in mind that not all the relays host VMs) such as a smart camera sending small size images to be processed.

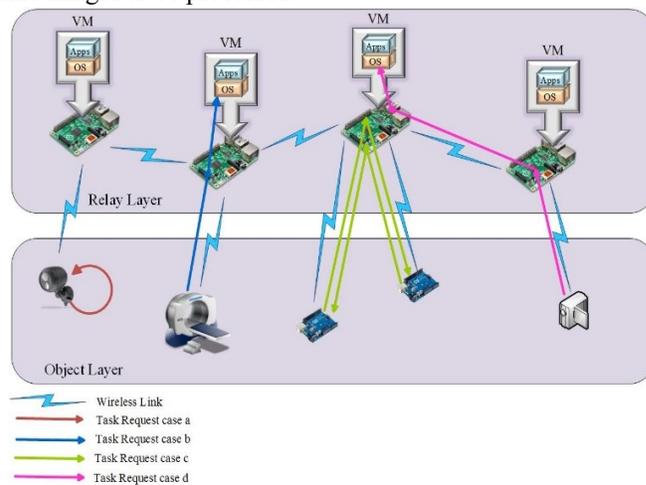

Fig. 1. The proposed architecture with P2P communication and processing

The MILP model objective consists of two main parts. The first part maximizes the number of logical end-to-end connections between objects and between both relays and other objects. Maximizing this number means maximizing the number of served tasks. The second part of the objective considers minimizing the total power consumption of all elements in our network. The total power consumption in our model is made up of two parts. The first part is the traffic induced power consumption in objects and relays caused by uplink and downlink traffic flow through the network. The uplink traffic is generated by the task requests (the row data) while the downlink traffic is the reduced traffic generated after task processing (the information). The second part of the power consumption equation represents the processing induced power consumption in objects and relays produced by the tasks processing in objects and hosted VMs.

The MILP model objective is subject to many constraints. These constraints are related to VMs placements, fairness constraint between the objects, tasks number limitations, uplink and downlink capacities, and processing capability limitations. For more clarity in the MILP expressions and notations, we have used superscripts to index the type of variables and parameters, while we have used subscripts as indices of these variables and parameters.

First, the sets, parameters, and variables of our P2P IoT MILP model are defined in Tables I and II:

**Table I List of parameters and their definitions**

| Notation | Description |
|---|---|
| $O$ | Set of objects |
| $VM$ | Set of virtual machines |
| $R$ | Set of relays |
| $P$ | Set of peers ($O \cup R$) |
| $TN$ | Set of all IoT network nodes ($TN = P \cup R$) |
| $P_r^R$ | Set of peers of relay $r$ |
| $N_a$ | Set of neighbors of node $a$ |
| $K$ | Set of tasks |
| $K^p$ | Subset of tasks that can be served by each peer p. $K^p \subset K$ |
| $R_p^P$ | Neighbor relay of peer $p$ if the peer is an object or $R_p^P$ is peer $p$ if the peer is the hosting relay |
| $Q_{ik}$ | Task $k$ required by object $i$ |
| $W_k$ | The workload required by each task $k$ (GHz) |
| $B^V$ | The number of possible locations occupied by VMs |
| $\psi_j^r$ | The processor capacity of each relay $j$ (GHz) |
| $\Omega_j^r$ | The maximum power consumed by the processor used in each relay $j$ (W) |
| $\psi_j^O$ | The processor capacity of each object $j$ ( (GHz) |
| $\Omega_j^O$ | The maximum power consumed by each object $j$ ( (W) |
| $M_k$ | The traffic demand of each task $k$ (bit/s) (row data) |
| $C_k$ | The traffic resulting after processing each task $k$ (bit/s) (information) |
| $L^{DO}$ | Maximum traffic that can be downloaded by each object (bits/s) |
| $L^{Dr}$ | Maximum traffic that can be downloaded by each relay (bits/s) |
| $L^{UO}$ | Maximum traffic that can be uploaded by each object (bits/s) |
| $L^{Ur}$ | Maximum traffic that can be uploaded by each relay (uploading tasks results) (bits/s) |
| $X_i$ | Maximum number of upload slots for each object $i$ |
| $E^{elec}$ | Energy consumed per bit by the electronics of the transceiver (Joules/bit) |
| $D_{mn}$ | Distance between any node pair in the IoT network $(m,n)$ (meter) $m, n \in TN$ |
| $\epsilon$ | Transmission amplifier power coefficient (Joule/bit.m$^2$) |

**Table II List of variables and their definitions**

| Notation | Description |
|---|---|
| $U_{ijk}$ | Binary variable which is set to 1 if peer $i$ processes task $k$ requested by object $j$, otherwise it is set to 0 |
| $V_j$ | Binary variable which is set to 1 if there is a virtual machine in that location otherwise it is set to 0 |
| $I_j^{DM}$ | Download rate (downloading task request) for each peer $j$ (kbps) |
| $I_i^{DC}$ | Download rate (downloading task result) for each object $i$ (kbps) |
| $I_i^{UM}$ | Upload rate (uploading task request) for each object $i$ (kbps) |
| $I_j^{UC}$ | Upload rate (uploading task result) for each peer $j$ (kbps) |
| $\lambda_{xy}^Q$ | Total traffic passing from relay $x$ (neighbor of source object) to relay $y$ (neighbor of destined object or hosting the destined VM) |
| $\lambda_{xy}^S$ | Total traffic (tasks result traffic) passing from the relay $x$ (neighbor of source object or source relay) to relay $y$ (neighbor of destined object) |
| $\lambda_{xyab}^Q$ | Relay to relay traffic $(x, y)$ passing through the link between the intermediate relays pair $(a, b)$ |
| $\lambda_{xyab}^S$ | Relay to relay traffic $(x, y)$ (tasks results traffic) passing through the link between the intermediate relays pair $(a, b)$ |
| $\lambda_{ab}^Q$ | Intermediate traffic between any two relays pair $(a, b)$ |
| $\lambda_{ab}^S$ | Intermediate traffic (tasks results traffic) between any two relays pair $(a, b)$ |

The total IoT network power consumption is composed of:
1. The processing induced power consumption of each peer, which can be calculated by summing the workloads of all processed tasks by the peer and multiplying the summation by the energy per processed bit. The processing power in our work is composed of two parts:
   a) Processing induced power consumption of each object:

$$P_j^{op} = \left( \sum_{i \in O, k \in K_j^P} U_{ijk} \cdot W_k \right) \cdot \frac{\Omega_j^o}{\psi_j^o} \quad (1)$$

$$\forall j \in O$$

b) Processing induced power consumption of each relay:

$$P_j^{rp} = \left( \sum_{i \in O, k \in K_j^P} U_{ijk} \cdot W_k \right) \cdot \frac{\Omega_j^r}{\psi_j^r} \quad (2)$$

$$\forall j \in R$$

2. The traffic induced power consumption, which consists of two basic parts, the sending part and the receiving part. Both parts are based on radio energy dissipation (Friis free-space equation) used in [18]. The power consumption is equal to the bit rate times the propagation energy per bit.

   a) The traffic induced power consumption of each object:

$$P_i^{otr} = \sum_{j \in P} \sum_{k \in K_j^P : i \neq j} U_{ijk} \cdot M_k \cdot \left( E^{elec} \right.$$

$$\left. + \epsilon \cdot D_{ig}^2 \right)$$

$$+ \left( \sum_{j \in O : i \neq j} \sum_{k \in KP_j} U_{jik} \cdot C_k \cdot \left( E^{elec} \right.\right.$$

$$\left.\left. + \epsilon \cdot D_{ig}^2 \right) \right)$$

$$+ \sum_{j \in O : i \neq j} \sum_{k \in K_j^P} U_{jik} \cdot M_k \cdot E^{elec}$$

$$+ \left( \sum_{j \in P : i \neq j} \sum_{k \in KP_i} U_{ijk} \cdot C_k \cdot E^{elec} \right) \quad (3)$$

$$g = R_i^P$$

$$\forall i \in O$$

The first two terms represent the sending power while the third and fourth parts represent the receiving power. The first term calculates the power consumed by each object in sending its requests to other peers in order to process them. The second part represents the power consumed by each object in sending back the results of the tasks processed by itself to the original request generator. The third part represents the power consumed by each object in receiving the task requests from other objects. The last part shows the power consumed by each object in receiving the results of its task requests.

b) Traffic induced power consumption of each relay:

$$P_a^{rtr} = \sum_{b \in N_a \cap R : a \neq b} \left( \lambda_{ab}^Q \cdot (E^{elec} \right.$$

$$\left. + \epsilon \cdot D_{ab}^2 ) \right)$$

$$+ \sum_{b \in N_a \cap R : a \neq b} \left( \lambda_{ab}^S \cdot (E^{elec} \right.$$

$$\left. + \epsilon \cdot D_{ab}^2 ) \right)$$

$$+ \sum_{j \in P_a^R \cap O} \left( I_j^{DM} \cdot (E^{elec} \right.$$

$$\left. + \epsilon \cdot D_{aj}^2 ) \right)$$

$$+ \sum_{i \in P_a^R \cap O} \left( I_i^{DC} \cdot (E^{elec} + \epsilon \cdot D_{ai}^2 ) \right)$$

$$+ \sum_{b \in N_a \cap R : a \neq b} \left( \lambda_{ba}^Q \cdot E^{elec} \right)$$

$$+ \sum_{b \in N_a \cap R : a \neq b} \left( \lambda_{ba}^S \cdot E^{elec} \right)$$

$$+ \sum_{i \in P_a^R \cap O} I_i^{UM} \cdot E^{elec} + \sum_{j \in P_a^R \cap O} I_j^{UC} \cdot E^{elec}$$

$$\forall a \in R \quad (4)$$

Traffic induced power consumption of the relays $P_a^{rtr}$ consists of 8 terms. The first four terms represent the sending power and the last four terms represent the receiving power. The first and second terms represent the power consumed in sending the task requests and task results respectively from a relay to another relay. The third and fourth terms calculate the power consumed in sending the task requests and task results respectively to the objects directly connected to that relay. The fifth and sixth terms describe the power consumed in receiving task requests and task results respectively by each relay from another neighbor relay. The seventh term calculates the power consumed by each relay in receiving the task requests from the directly connected object while the last term represents the power consumed by each relay in receiving the task results from other peers (directly connected object to the relay or relay hosting

VM).

**Objective**: Maximize

$$\left(\sum_{i\in O, j\in P, k\in K_j^P} F \cdot U_{ijk}\right) - \left(\sum_{j\in O} P_j^{op} + \sum_{j\in R} P_j^{rp}\right)$$

$$- \left(\sum_{i\in O}(P_i^{otr}) + \sum_{a\in R}(P_a^{rtr})\right) \quad (5)$$

Equation (5) gives the model objective where the number of logical end-to-end connections between objects and other peers is maximized while the network power consumption and the processing power consumption are minimized. The parameter $F$ takes care of the units and is also used to scale the number of connections so that they become comparable in magnitude to the consumed power.

Constraints
**Subject to:**
1. U indicator setting constraints

$$\sum_{j\in P} U_{ijk} \le Q_{ik} \quad (6)$$
$$\forall i \in O, \forall k \in K$$

Constraint (6) ensures that only one peer (one object or one relay) can serve each request of each object.

2. Fairness constraints

$$\sum_{k\in K_j^P} U_{ijk} = \sum_{k\in K_i^P} U_{jik} \quad (7)$$
$$\forall i \in O, \forall j \in O$$

Constraint (7) is the fairness constraint which ensures that each object reciprocates equally to other objects that serve a request of this object.

3. Virtual Machine Calculations constraints

$$\sum_{i\in O, k\in K_j^P} U_{ijk} \ge V_j \quad (8)$$
$$\forall j \in VM$$

$$\sum_{i\in O, k\in K_j^P} U_{ijk} \le A \cdot V_j \quad (9)$$
$$\forall j \in VM$$

$$\sum_{j\in VM} V_j = B^v \quad (10)$$

Constraints (8) and (9) locate a virtual machine in an appropriate relay in order to process the requested tasks. Constraint (10) limits the number of selected locations occupied by the virtual machines to 10 only out of 25 possible locations.

4. Processing power consumption calculations

$$\sum_{i\in O, k\in K_j^P} U_{ijk} \cdot W_k \le \psi_j^O \quad (11)$$
$$\forall j \in O$$

$$\sum_{i\in O, k\in K_j^P} U_{ijk} \cdot W_k \le \psi_j^r \quad (12)$$
$$\forall j \in R$$

Constraints (11) and (12) ensure that the summation of the whole workloads of processed tasks by each object and each relay respectively do not exceed its maximum processing workload capability

5. Traffic calculations and capacity constraints

$$\lambda_{xy}^Q = \left(\left(\sum_{i\in P_x^R \cap O}\left(\sum_{j\in P_y^R: i\ne j}\left(\sum_{k\in K_j^P} U_{ijk} \cdot M_k\right)\right)\right)\right) \quad (13)$$
$$\forall x \in R, \forall y \in R: x \ne y$$

$$\lambda_{xy}^S = \left(\left(\sum_{i\in P_x^R}\left(\sum_{j\in P_y^R \cap O: i\ne j}\left(\sum_{k\in K_j^P} U_{ijk} \cdot C_k\right)\right)\right)\right) \quad (14)$$
$$\forall x \in R, \forall y \in R: x \ne y$$

$$\sum_{b\in N_a\cap R: a\ne b}\lambda_{xyab}^Q - \sum_{b\in N_a\cap R: a\ne b}\lambda_{xyba}^Q$$
$$= \begin{cases} \lambda_{xy}^Q & \text{if } a = x \\ -\lambda_{xy}^Q & \text{if } a = y \\ 0 & \text{otherwise} \end{cases} \quad (15)$$
$$\forall x \in R, \forall y \in R, \forall a \in R: x \ne y$$

$$\sum_{b\in N_a\cap R: a\ne b}\lambda_{xyab}^S - \sum_{b\in N_a\cap R: a\ne b}\lambda_{xyba}^S$$
$$= \begin{cases} \lambda_{xy}^S & \text{if } a = x \\ -\lambda_{xy}^S & \text{if } a = y \\ 0 & \text{otherwise} \end{cases} \quad (16)$$
$$\forall x \in R, \forall y \in R, \forall a \in R: x \ne y$$

$$\lambda_{ab}^Q = \sum_{x\in R}\sum_{y\in R: x\ne y}\lambda_{xyab}^Q \quad (17)$$
$$\forall a \in R, b \in N_a \cap R: a \ne b$$

$$\lambda_{ab}^S = \sum_{x\in R}\sum_{y\in R: x\ne y}\lambda_{xyab}^S \quad (18)$$
$$\forall a \in R, b \in N_a \cap R: a \ne b$$

Constraints (13) and (14) calculate the transient traffic between relays due to P2P traffic (task requests and the results traffic). Constraints (15) and (16) represent the flow conservation of the traffic between the source relay (requester's (object) neighbor) and the destination relay (serving peer's neighbor or host) through the intermediate relays. Constraints (17) and (18) calculate the traffic flows through each intermediate relay.

$$I_j^{DM} = \sum_{i \in O, k \in K_j^P : i \neq j} U_{ijk} \cdot M_k \quad (19)$$
$$\forall j \in P$$

$$I_i^{DC} = \sum_{j \in P, k \in K_j^P : i \neq j} U_{ijk} \cdot C_k \quad (20)$$
$$\forall i \in O$$

$$I_j^{DM} \leq L^{DO} \quad (21)$$
$$\forall j \in O$$

$$I_i^{DC} \leq L^{DO} \quad (22)$$
$$\forall i \in O$$

$$I_j^{DM} \leq L^{DR} \quad (23)$$
$$\forall j \in R$$

Constraint (19) calculates the download rate of each peer by summing the received traffic demand of each requested task from other objects selected to serve them. Constraint (20) calculates the download rate of the reduced traffic (resulting information) received by each object. Constraints (21), and (22) limit the download rate of each object to its maximum value, while constraint (23) limits the download rate of each relay to its maximum value.

$$I_i^{UM} = \sum_{j \in P, k \in K_j^P : i \neq j} U_{ijk} \cdot M_k \quad (24)$$
$$\forall i \in O$$

$$I_j^{UC} = \sum_{i \in O, k \in K_j^P : i \neq j} U_{ijk} \cdot C_k \quad (25)$$
$$\forall j \in P$$

$$I_i^{UM} \leq L^{UO} \quad (26)$$
$$\forall i \in O$$

$$I_j^{UC} \leq L^{UO} \quad (27)$$
$$\forall j \in O$$

$$I_j^{UC} \leq L^{UR} \quad (28)$$
$$\forall j \in R$$

$$\sum_{j \in P, k \in K_j^P : i \neq j} U_{ijk} \leq X_i \quad (29)$$
$$\forall i \in O$$

$$\sum_{i \in O, k \in K_j^P : i \neq j} U_{ijk} \leq X_i \quad (30)$$
$$\forall j \in O$$

Constraint (24) calculates the upload rate of each object by summing the uploaded task traffic demands. While constraint (25) calculates the upload rate of each peer that results from sending the reduced traffic (the resulting information from task processing). Constraints (26), and (27) limit the upload rate of each object to its maximum value while constraint (28) limits the upload rate of each relay to its maximum value. Constraints (29) and (30) limit the number of upload slots of each object.

III. MILP MODEL EVALUATION AND RESULTS

Our IoT nodes, depicted in Fig. 1, consist of 25 objects and 25 relays distributed over an area of 30m × 30m [19]. The objects are distributed randomly while the relays are distributed uniformly, every 6m as shown in Fig. 2. All devices in the IoT network communicate using the Zigbee protocol.

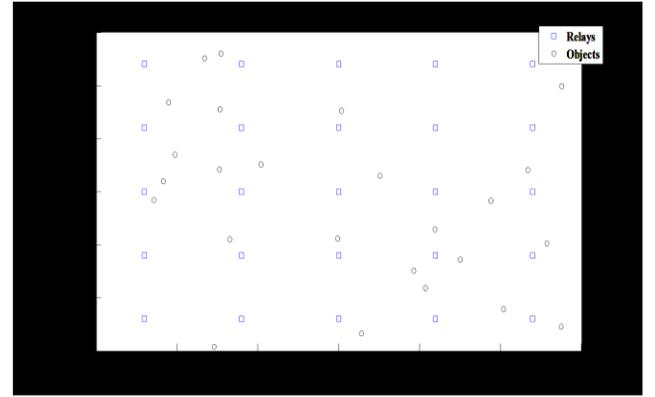

Fig. 2 IoT distribution in space

Table III lists the model input parameters. We have used the Arduino 101 as an IoT object as it is one of the most power efficient processors with a higher clock speed compared to other types of Arduino [20]. Arduino 101 is referred to as Genuino 101 outside USA [21]. We used the Raspberry pi 3 in the relays, with processing capability of 1.2 GHz [22]. We assumed the traffic demand of the first task is 250 bit/s representing applications with small traffic volume in the range 0-250 bit/s. We assumed other values of traffic close to this one in a consistent way to comply with the link capacity limit constraint and to be very close to practical IoT applications. The data rates thus considered were 240b/s representing a heartbeat sensor and 2.4 kb/s associated with blood glucose level sensors and temperature readings [23]. The range of traffic values considered resulted in heterogeneous tasks that have to be tackled by our optimization model [24], [25]. In Bit Torrent, the typical value for the maximum number of upload slots for each peer is 4 [26]. We have considered, in our P2P communication system, a range of different numbers of upload slots from 1 to 10 slots per object. We found that the average value of upload slots that ensures the highest percentage of executed tasks is 4.

TABLE III MILP MODEL INPUT PARAMETERS

| Parameter Description | value |
|---|---|
| Energy per bit consumed by the electronics of the transceiver ($E^{elec}$) | 50 nJ/bit [27] |
| Maximum download rate of each peer (objects and relays) ($L^{DO}$ and $L^{DR}$) | 10 kb/s, 25 kb/s [24,28] |
| Maximum upload rate of each (objects and relays) ($L^{UO}$ & $L^{UR}$) | 5 kb/s, 25 kb/s [28,29] |
| The processor capacity of object ($\psi_j^O$) | 32 MHz [20] |
| The processor Capacity of relay ($\psi_j^R$) | 1.2 GHz [22] |
| CPU maximum power consumption in objects ($\Omega_j^O$) | 347 mW [20] |
| CPU maximum power consumption in relay ($\Omega_j^R$) | 3.7 W [30] |
| Transmission amplifier power coefficient ($\epsilon$) | 255 pJ/bit.$m^2$ [27] |
| The requested workload for each task ($W_k$), $k \in K$ | 0.01 GHz, 0.012 GHz, 0.015 Hz, 0.02 GHz, 0.05 GHz, 0.1 GHz, 0.2 GHz, 0.3 GHz, 0.4 GHz, 0.5 GHz [31] |
| Traffic generated by each task request ($M_k$), $k \in K$ | 250b/s, 500b/s, 750b/s, 1000b/s, 1250b/s, 1750b/s, 2000b/s, 2250b/s, 2500b/s, 2750b/s [23-25] |
| Traffic generated by each task result after reduction ($C_k$), $k \in K$ | 25b/s, 100b/s, 225b/s, 400b/s, 625b/s, 1050b/s, 1400b/s, 1800b/s, 2125b/s, 2475b/s |
| Maximum number of upload slots for each peer ($X_i$) | 4 [26] |
| IoT nodes distribution area | 30m × 30m [19] |
| Range of task weight ($F$) for all scenarios | {0, 0.1, 0.2, 0.3, 0.6, 0.9, 1.2, 1.5, 1.8} |
| Scale factor with large value (M) | 1000000 |

As alluded earlier, we have considered three scenarios. The first scenario is the relays only scenario. This restricts the processing of all requested tasks to 10 VMs out of 25 possible locations. This scenario is implemented by setting the number of end-to-end connections between the objects to zero, to ensure that no objects respond to any task request, i.e. equation (31):

$$\sum_{j \in O, k \in K_j^P} U_{ijk} = 0 \quad (31)$$
$$\forall\, i \in O$$

The second scenario is the objects only scenario which restricts the processing of the requested tasks by the IoT to objects only. This scenario is implemented by setting the total number of VMs to zero (it actually means the number of relays hosting VMs equals zero), i.e. equation (32):

$$\sum_{j \in VM} V_j = 0 \quad (32)$$

The last scenario allows cooperation between the relays hosting VMs and the objects in order to process the requested tasks. Fig. 3 shows the processing induced power consumption of the three scenarios. The $x$ axis represents the range of different values of task weights $F$ multiplied by the variable U as shown in (5) (the objective function). This range is used to scale the number of connections to be comparable to the amount of consumed power.

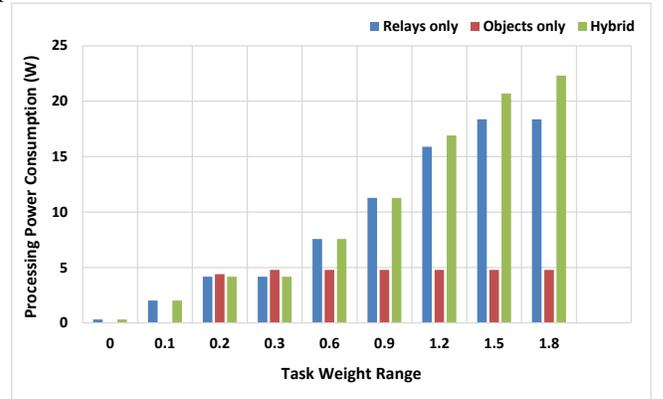

Fig. 3. Total processing induced power consumption in the three scenarios

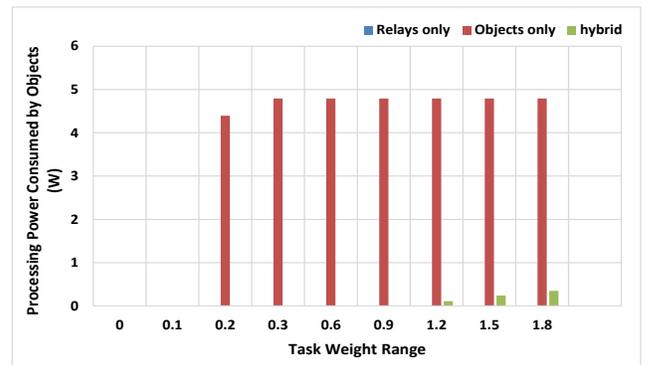

Fig. 4. Processing induced power consumption by objects in the three scenarios

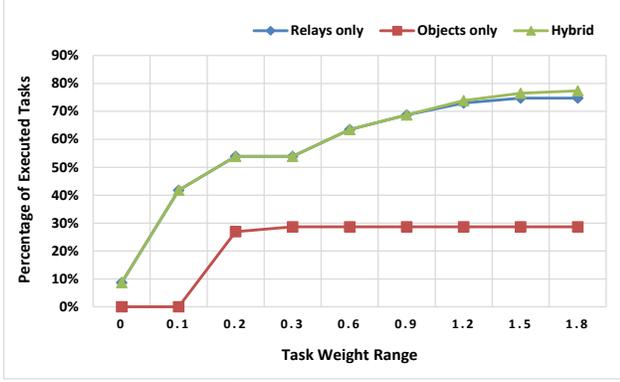

Fig. 5 percentage of executed tasks by the three scenarios

Fig. 3 shows that the hybrid and relays only scenarios consume the same amount of processing induced power at task weight values in the range ($F=0 \sim 0.9$) as there are no tasks executed by the objects in the hybrid scenario at these values as shown in Fig. 4. The inefficiency of the objects-processors used and the effect of the power optimization at such low scale factor result in blocking the requested tasks instead of implementing them by the objects. The power inefficiency of a processor used in object processing only is clearly illustrated in Fig. 5 at task weight values ($F= 0.2$ and $F= 0.3$). At these values, the objects only scenario executes less tasks than the other two scenarios (about half), but it consumes more processing power than both of them as shown in Fig. 3.

Starting at $F=0.3$ and up to the highest task weight, the objects only scenario consumes the same amount of processing induced power. The low utilization of the P2P layer in the objects only scenario is attributed to two reasons. Firstly, the effect of the fairness constraint and secondly, the most effective reason is the low capacity of the processors in the IoT objects. This low capacity is clearly seen in Table IV as the objects in the objects only scenario drop tasks (5 to 10) as the workloads of these tasks are larger than the processor capacity of the objects ($\psi_j^o$) as shown in Table III.

A general trend followed by both hybrid and the relays only scenarios towards higher power consumption for higher task weights can be seen in Fig. 3. Starting from task weight $F=1.2$ a small gap is observed between the two scenarios and this grows as the task weight increases. This gap is caused by the higher power consumption of the hybrid scenario compared to the relays only scenario because of the internal processing of the objects in the hybrid scenario. Due to the limited number of upload slots available for each object, an object tends to process its requests internally instead of using the free upload slots. Accordingly, internal processing allows the objects to send more task requests with higher workloads to relays to be processed. Therefore, the relays in the hybrid scenario consume more processing induced power than the relays in the relays only scenario as shown in Fig. 6.

To clarify, we consider task $k_9$ in Table IV as an example. In the hybrid scenario, task $k_9$ is requested by objects 8, 15 and 25 and in the relays only scenario are only requested by objects 15 and 25 but not by 8. This means that the request by object no. 8 is blocked. Therefore, by checking object 8, we notice:

1. Object 8 in the hybrid scenario processes internally task request $k_2$ and sends $k_1$, $k_8$ and $k_9$ task requests to other peers. The total generated traffic as a result of sending all these requests is 5000 b/s which is the maximum limit of the upload capacity of each object (the traffic generated by each task request is illustrated in Table III).
2. Obviously, in the relays only scenario, internal processing is not allowed, therefore object 8 sends requests $k_1$, $k_2$ and $k_8$ to relays hosting VMs while task request $k_9$ is blocked. The total upload traffic due to requests is 3000 b/s which leaves only 2000 b/s of allowed traffic that can be uploaded by object 8. This (ie 2000 b/s) is not enough to transmit $k_9$ and that results in blocking this request instead of sending it to be served. In addition, blocking $k_9$ by object 8 in particular is due to the power optimization and its impact on the behaviour of the object. Since the object tries to send tasks with the lowest processor workload and lowest traffic demand requirements to be served by other peers, this results in blocking $k_9$.

Table IV
Tasks execution map at task weight ($F=1.8$)

| Task ID | Total No. of Task Requests | Total No. of Served Tasks | | |
| --- | --- | --- | --- | --- |
| | | Objects Only Scenario | relays Only Scenario | Hybrid Scenario |
| $k_1$ | 15 | 12 | 15 | 15 |
| $k_2$ | 10 | 6 | 10 | 10 |
| $k_3$ | 15 | 10 | 15 | 15 |
| $k_4$ | 8 | 5 | 8 | 8 |
| $k_5$ | 14 | 0 | 14 | 14 |
| $k_6$ | 11 | 0 | 11 | 11 |
| $k_7$ | 9 | 0 | 6 | 6 |
| $k_8$ | 13 | 0 | 5 | 6 |
| $k_9$ | 11 | 0 | 2 | 3 |
| $k_{10}$ | 9 | 0 | 0 | 1 |

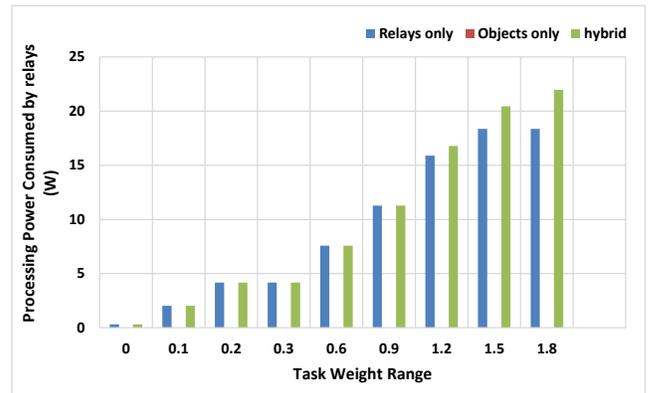

Fig. 6. Processing induced power consumption by relays in three scenarios

As a result of the power optimization, there is a general pattern followed by the objects in our network when they send their requests to be served by other peers. First, to make sure that the objects requests are satisfied using the lowest processing and network power consumption, objects search for the nearest available relays hosting VMs starting with ones that are directly connected to the object (objects' neighbors) then the circle of search is increased to include other relays starting from the nearest to the furthest. The implication is that the results in Fig. 7 show that the traffic induced power consumed by relays is more than the power used by the objects. This difference increases with increase in the task weight in both hybrid and relays only scenarios. In the hybrid scenario, when the model starts serving more tasks than the relays hosting VMs can handle because of traffic and processing capacity constraints, the objects serve tasks using their own processors (internal processing) as shown in Fig. 4. This starts at $F$=1.2 and continues beyond. Given that it is internal processing, it is of interest to understand the drivers behind the increase in the traffic induced power consumption in the relays. In this scenario, the internal processing affects the relays behavior resulting in serving more tasks with higher workload. Sending task requests with high workloads to relays hosting VMs results in consuming more traffic induced power by relays. In the objects only scenario, the objects either serve task requests using their own resources if they able to, or send the requested tasks to other objects to be served. Sending tasks to other objects while satisfying the fairness constraint can lead to sending the requests to remote objects. This results in higher network power consumption in the relays. Consequently, the traffic induced power consumption in the relays in the objects only scenario is higher than the power consumption in the objects only scenario. It is even higher than the power consumption in the relays in other scenarios as illustrated in Fig. 7 at task weight value $F$=0.3. However, as discussed earlier, the low capacity of the processor used in IoT objects results in low and constant serving tasks rate for other values of task weight range. This leads in turn to a constant consumption of traffic induced power for all devices in our network in the objects only scenario.

After considering the processing and traffic induced power consumption of the three scenarios, it is clearly seen that the hybrid scenario consumes the highest amount of total power compared to the other scenarios. Moreover, the relays only scenario consumes a comparable power with 8% power saving. Finally, the objects only scenario has the least total power consumption with power saving up to 62% compared to the hybrid scenario.

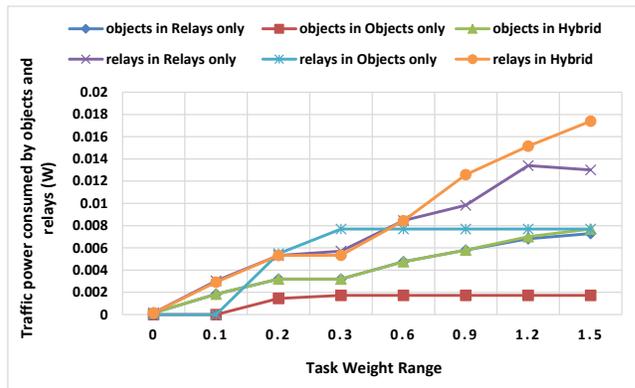

Fig. 7. Traffic induced power consumption by objects and relays in three scenarios

## IV. ENERGY EFFICIENT P2P IOT NETWORKS HEURISTIC AND RESULTS

In this section we try to mimic the behavior of our energy efficient P2P IoT MILP model in real time by developing a P2P energy efficient IoT task processing and traffic routing heuristic. The EEVIPN heuristic is illustrated in Fig. 5.8. It considers the hybrid scenario as it is the generic scenario that can be used to build other scenarios such as the relays only and the object only scenarios. To determine the total power consumption (TPC), the heuristic determines the type and the optimum place of the peer to be used to serve the processing tasks according to the serving constraints of each peer. The serving constraints can be summarized as follows:

i. The processing task should not have been served by any other peer before.
ii. The upload traffic of each candidate peer should not exceed the maximum limit.
iii. The download traffic of each candidate peer should not exceed the maximum limit.
iv. The upload slots of each object should not exceed the specified maximum number.
v. The number of candidate relays hosting VMs should not exceed the specified maximum number of serving relays.
vi. There should be sufficient processor capacity in each candidate peer to accommodate the processing task workload.

Recall that these are the general serving constraints and could be changed according to the type of the serving peer. For example, if the candidate peer is a relay then all the serving constraints should be considered. If the candidate peer is an object (not the task requester), constraint (v) will not be applied. For the internal processing scenario, the heuristic should check constraints (i) and (vi) because the requested task is served by the requester object internally and as a result, there will be no external data processing neither traffic flow.

For each task requested by an object, the heuristic first checks all the candidate relays hosting VMs in the network. Starting with relays, is an attempt by the heuristic to mimic the MILP model behavior at the lowest values of task weight, by looking for candidate relays as serving peers. The heuristic first checks relays hosting VMs due to the power efficiency of their processors compared to the power efficiency of the objects only processors. It also checks the relays first due to their high ability to serve all types of requested processing tasks. The serving constraints of the first candidate relay are investigated by the heuristic. If all these constraints are met, then the link between the requester and the serving relay is set. The requested task is served and the processing power $P_j^{Rp}$ of each relay is calculated. The heuristic loops for the rest of the relays hosting VMs for all requested tasks by all objects. It finally calculates all the processing induced power of all serving relays. If the requested task is served by an object, there are two cases, the first case represents internal processing. In the second case, the object serves another object. In this case, the Tit-for-Tat constraint (the fairness constraint, equation (7)) should be applied to guarantee equal reciprocity between the two objects intending to serve each other. In both cases, if all serving constraints are met then the link between the requester and the serving object is set. The candidate object serves the requested processing task and the processing induced power consumed by the object-processor $P_j^{op}$ is calculated. After checking all the possible serving peers for all requested tasks by all requesting objects, the traffic induced power consumption of each object $P_i^{otr}$ is calculated. In addition, the power consumption of each relay $P_a^{rtr}$ caused by cross traffic between the requesting objects and the serving peers is calculated. The traffic induced power consumption of each relay is composed of two basic parts. The first represents the power consumption due to traffic flowing between relays. The heuristic tries to route the traffic between node $x$ (the directly connected relay to the requesting object) and node $y$ (the directly connected relay to the serving object or hosting the serving VM) by using a minimum hop algorithm in order to minimize the traffic induced power consumption of each relay. The other part of $P_a^{rtr}$ is the network power consumption due to the traffic flowing between relays and the request generator and serving objects. Finally, the heuristic calculates the number of served tasks by all peers $NST$ and the total power consumption $TPC$.

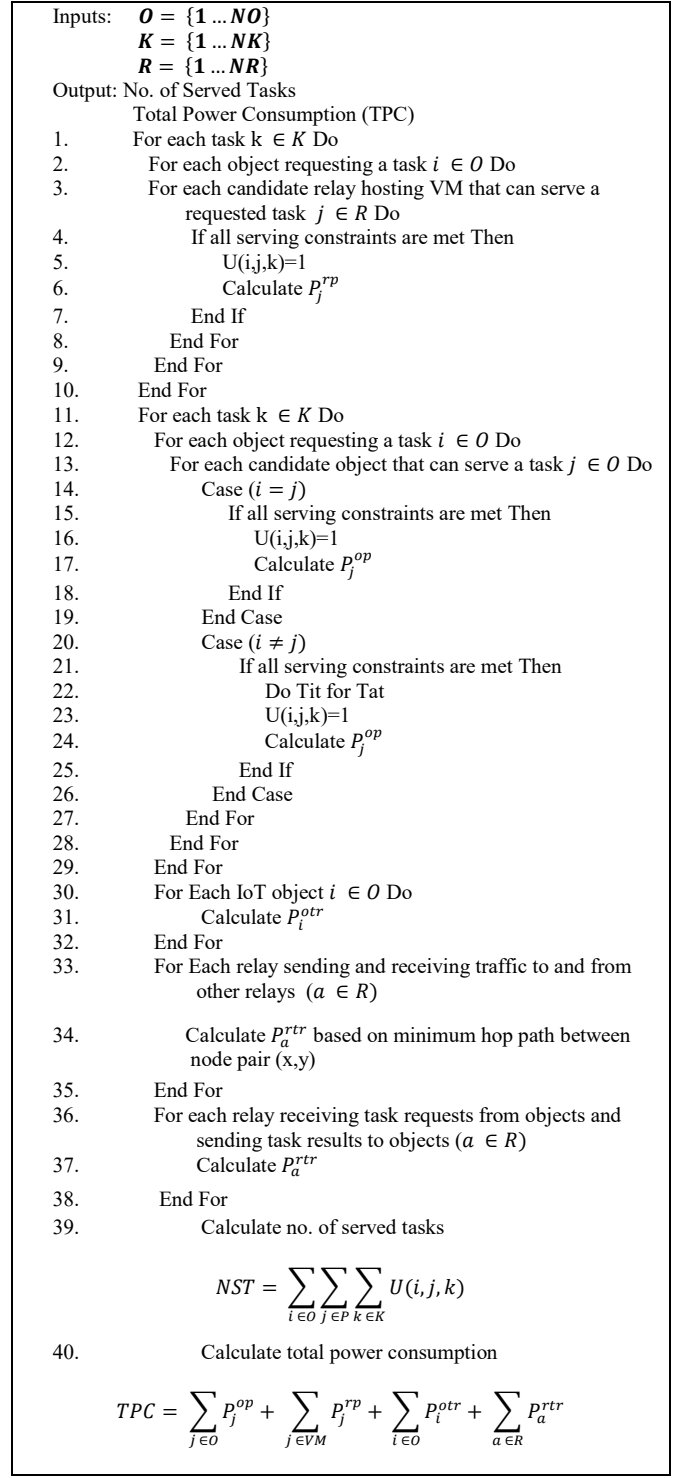

Fig. 8. P2P IoT heuristic

Fig. 9 presents the total power consumption of both MILP and EEVIPN heuristic versus the percentage of served tasks. It is clearly seen that the power consumption of the MILP and EEVIPN heuristic are comparable. The highest percentage of served tasks that can be achieved is 77% by the hybrid scenario in the MILP model. Therefore, we do not show results beyond 80% of served tasks as these cases will consume the same amount of power. It should be noted that in the hybrid scenario the MILP model consumes higher power than the heuristic when serving higher than 70% of the requested tasks because of the higher VMs utilization as clearly shown in Fig. 11. The higher utilization of VMs results from the internal processing by the objects at higher percentage of tasks execution as mentioned before in the discussion of the results in Fig. 3. There are no tasks served by the objects in the hybrid scenario in the heuristic as illustrated by Fig. 10. In the hybrid scenario, tasks with small workloads are served by relays as the heuristic starts task assignment with relays. After that, the heuristic tends to assign the remaining tasks (unserved) to objects where the tasks have workload requirements higher than the objects capabilities. As such, objects are not exploited in this scenario. Moreover, this results in both the hybrid scenario and relays only scenario (in heuristic) following the same behavior in executing tasks. This results in the two scenarios consuming the same amount of power as clearly shown in Fig. 9. Fig. 9 shows that the objects only scenario (heuristic) consumes higher power than MILP model. This small difference is attributed to the impact of the network power consumption and specifically the power consumed by the relays as shown in Fig. 12. In the MILP model (objects only scenario), if the tasks are not served internally by the objects then the model optimizes the choice of the serving objects according to the fairness constraint in addition to the distances from the requesting objects to the serving objects in order to reduce the power consumption. In the heuristic, the search for serving objects is carried out sequentially regardless of their locations. This results in the relays consuming more power especially in cases where the tasks are sent to remote serving objects. A similar observation can be made about the difference between the power consumed by relays (due to traffic) in both hybrid and relays only scenarios. In the heuristic, the relays consume higher traffic induced power than in the MILP. This is similar to the objects only scenario. It is also caused by sending the requests far apart in order to be served by the candidate serving relays.

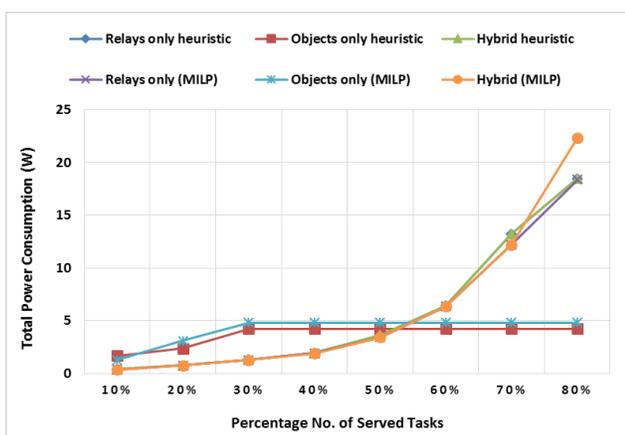

Fig. 9. Total power consumption evaluated using heuristic and MILP model

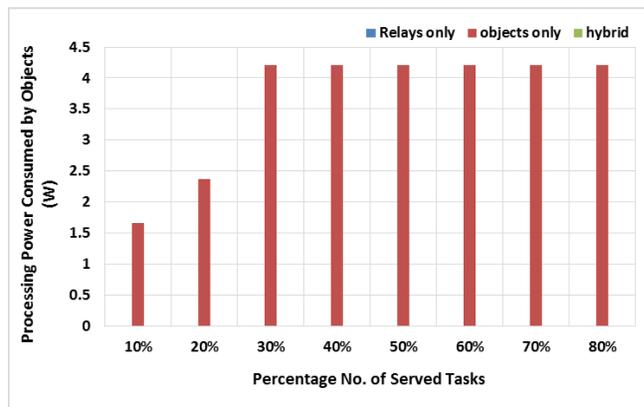

Fig. 10. Processing induced power consumption of objects

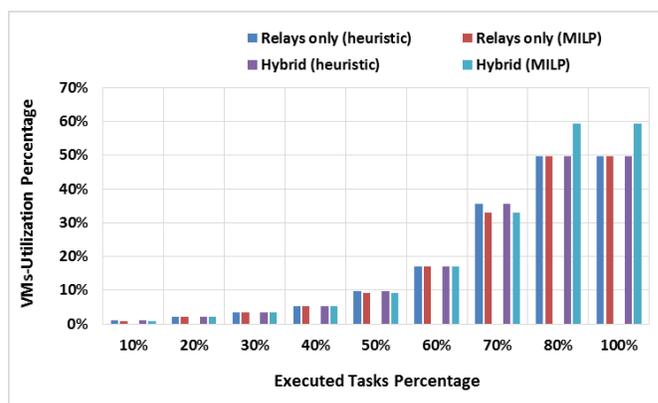

Fig 11. VMs- utilization in hybrid and relays only scenarios

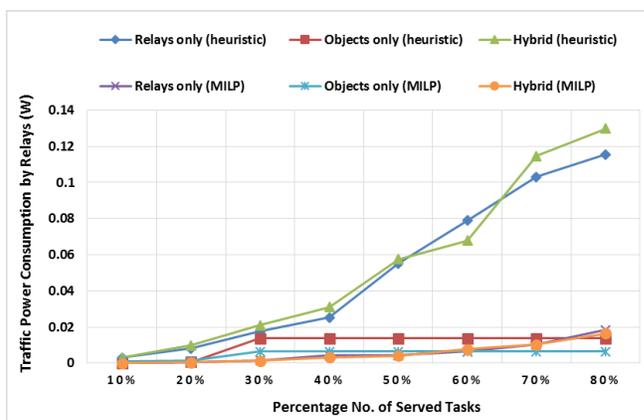

Fig. 12. Traffic induced power consumption of Relays

V. CONCLUSIONS

In this paper, we have investigated the energy efficiency of an IoT virtualization framework with P2P network and edge computing. This investigation has been carried out by considering three different scenarios. A MILP was developed to maximize the number of processing tasks served by peers and minimize the total power consumption of the network.

Our results show that the hybrid scenario serves up to 77% (57% on average) processed task requests, but with higher energy consumption compared with other scenarios. The relays only scenario can serve 74% (57% on average) of the processing task requests with 8% of power saving and 28% (22% on average) of task requests can be successfully handled by applying the objects only scenario with 62% power saving.

The results also revealed the low percentage of addressed task requests in the objects only scenario resulting from the capacity limit of the IoT objects' processors. In addition, the small difference between the serving percentage of hybrid scenario and relays only scenario resulted from the allowed internal processing of objects in the hybrid scenario.

For real time implementation, we have developed the EEVIPN heuristic based on the MILP model concepts. The heuristic achieved a comparable power efficiency and comparable number of executed tasks to the MILP model. The hybrid Scenario in the heuristic executes up to 74% of the total tasks (MILP 77%), up to 74% of tasks by the relays only scenario (MILP 74%) while the objects only scenario executes up to 21% of the tasks (MILP 28%).

ACKNOWLEDGMENTS

We would like to acknowledge funding from the Engineering and Physical Sciences Research Council (EPSRC) for the INTERNET (EP/H040536/1) and STAR (EP/K016873/1) projects. The first author Mrs. Zaineb Al-Azez would like to thank Mr. Ahmed Al-Quzweeni for many helpful discussions and the Higher Committee for Education Development in Iraq (HCED) for funding her PhD scholarship. All data are provided in full in the results section of this paper.

REFERENCES

[1] F. Ganz, D. Puschmann, P. Barnaghi and F. Carrez, "A practical evaluation of information processing and abstraction techniques for the Internet of Things," in *IEEE Internet of Things Journal*, vol. 2, no. 4, pp. 340-354, Aug. 2015.
[2] J. Pan, and J. McElhannon, "Future Edge Cloud and Edge Computing for Internet of Things Applications", in *IEEE Internet of Things Journal*, vol. 5, no. 1, pp. 439-449, Feb. 2018.
[3] A. Al-Fuqaha, M. Guizani, M. Mohammadi, M. Aledhari and M. Ayyash, "Internet of Things: A survey on enabling technologies, protocols, and applications," in *IEEE Communications Surveys & Tutorials*, vol. 17, no. 4, pp. 2347-2376, Fourthquarter 2015.
[4] S. H. Shah and I. Yaqoob, "A survey: Internet of Things (IOT) technologies, applications and challenges," 2016 IEEE Smart Energy Grid Engineering (SEGE), Oshawa, ON, 2016, pp. 381-385.
[5] L. Nonde, T. E. H. El-Gorashi, and J. M. H. Elmirghani: Energy efficient virtual network embedding for cloud networks, *IEEE/OSA j. Lightwave technology*, Vol. 33. No. 9. pp. 1828-1849, 2015.
[6] A.Q. Lawey, T. E. H. El-Gorashi and J. M. H. Elmirghani: BitTorrent content distribution in optical networks, *IEEE/OSA j. Lightwave technology*, Vol. 32. No. 21. pp. 4209-4225, 2014.
[7] X. Dong, T. E. H. El-Gorashi and J. M. H. Elmirghani: Green IP over WDM networks with data centers, *IEEE/OSA j. Lightwave technol.*, Vol. 29. No. 12. pp. 1861-1880, 2011.
[8] X. Dong, T. E. H. El-Gorashi and J. M. H. Elmirghani: On the energy efficiency of physical topology design for IP over WDM networks, *IEEE/OSA j. Lightwave technol.*, Vol. 30. No. 12. pp. 1931-1942, 2012.
[9] M. Chiang and T. Zhang, "Fog and IoT: an overview of research opportunities," in *IEEE Internet of Things Journal*, vol. 3, no. 6, pp. 854-864, Dec. 2016.
[10] N. I. Osman, T. E. H. El-Gorashi, L. Krug, J. M. H. Elmirghani, "Energy-efficient future high-definition TV", *IEEE/OSA Journal of Lightwave Technology*, vol. 32, no. 13, pp. 2364-2381, 2014.
[11] A.Q. Lawey, T. E. H. El-Gorashi and J. M. H. Elmirghani: Distributed energy efficient clouds over core networks, *IEEE/OSA j. Lightwave technol.*, Vol. 32. No. 7. pp. 1261-1281, 2014.
[12] A. M. Al-Salim, A. Q. Lawey, T. E. H. El-Gorashi, J. M. H. Elmirghani, "Energy efficient big data networks: Impact of volume and variety", *IEEE Transactions on Network and Service Management*, no. 99, pp. 1-1, 2017.
[13] Z. T. Al-Azez, A. Q. Lawey, T. E. H. El-Gorashi and J. M. H. Elmirghani, "Virtualization framework for energy efficient IoT networks," *2015 IEEE 4th International Conference on Cloud Networking (CloudNet)*, Niagara Falls, ON, 2015, pp. 74-77.
[14] Z. T. Al-Azez, A. Q. Lawey, T. E. H. El-Gorashi and J. M. H. Elmirghani, "Energy efficient IoT virtualization framework with passive optical access networks," *2016 18th International Conference on Transparent Optical Networks (ICTON)*, Trento, 2016, pp. 1-4.
[15] F. Jalali, A. Vishwanath, J. de Hoog and F. Suits, "Interconnecting Fog computing and microgrids for greening IoT," *2016 IEEE Innovative Smart Grid Technologies - Asia (ISGT-Asia)*, Melbourne, VIC, 2016, pp. 693-698.
[16] Morabito and N. Beijar, "Enabling data processing at the network edge through lightweight virtualization technologies," *2016 IEEE International Conference on Sensing, Communication and Networking (SECON Workshops)*, London, 2016, pp. 1-6.
[17] U. Urosevic and Z. Veljovic, "Distributed MIMO solutions for peer-to-peer communications in future wireless systems," *2016 24th Telecommunications Forum (TELFOR)*, Belgrade, 2016, pp. 1-4.
[18] [31] J. Huang, Y. Meng, X. Gong, Y. Liu, and Q. Duan, "A novel deployment scheme for green Internet of Things," *IEEE Internet Things J*, vol. 1, pp. 196-205, 2014.
[19] P. Szczytowski, A. Khelil, and N. Suri, "DKM: Distributed k-connectivity maintenance in wireless sensor networks," in Proc. IEEE 9th Annu. Conf. Wireless On-Demand Net. Syst. Services (WONS), Courmayeur, Italy, Jan. 2012, pp. 83–90.
[20] THOMAS LEXTRAIT, Arduino: Power Consumption Compared [online]. Available: https://tlextrait.svbtle.com/arduino-power-consumption-compared
[21] Arduino and Genuino products [online]. Available: https://www.arduino.cc/en/Main/ArduinoBoard101.
[22] Raspberry Pi3 Model B Technical Specification [online]. http://docs-europe.electrocomponents.com/webdocs/14ba/0900766b814ba5fd.pdf
[23] N.Javaid, S.Faisal, Z.A.Khan, D.Nayab, M.Zahid, "Measuring Fatigue of Soldiers in Wireless Body Area Sensor Networks", *8th IEEE International Conference on Broadband and Wireless Computing, Communication and Applications (BWCCA'13),* Compiegne, France, 2013, pp. 227-231.
[24] W. Xu, W. Liang, X. Jia, Z. Xu, "Maximizing sensor lifetime in a rechargeable sensor network via partial energy charging on sensors", *Proc. 13th IEEE Int. Conf. Sensing Commun. Netw.*, 2016.
[25] C. Tung et al., "A mobility enabled inpatient monitoring system using a zigbee medical sensor network", *Sensors*, vol. 14, no. 2, pp. 2397-2416, 2014.
[26] A. Q. Lawey, T. El-Gorashi and J. M. H. Elmirghani, "Impact of peers behavior on the energy efficiency of BitTorrent over optical networks," *2012 14th International Conference on Transparent Optical Networks (ICTON)*, Coventry, 2012, pp. 1-8.
[27] Frank Comeau, Nauman Aslam, "Analysis of LEACH Energy Parameters", Procedia Computer Science, Volume 5, 2011, pp. 933-938.
[28] Cisco 910 Industrial Router [online]. Available: https://www.cisco.com/c/en/us/support/routers/910-industrial-router/model.html
[29] Intel Atom Z510 specifications [online]. Available: http://www.cpu-world.com/CPUs/Atom/IntelAtom%20Z510%20AC80566UC005DE.html
[30] SENSORO Alpha Product Suite- Alpha Node-4AA [online]. Available: https://www.sensoro.com/static/node4aa_en.pdf [last accessed: 11 Jun 2018]
[31] Ren, Zujie, et al., "Workload Characterization on a Cloud Platform: An Early Experience." *International Journal of Grid and Distributed Computing*, vol. 9. No. 6, 2016, pp. 259-268.